\documentclass[acmsmall]{acmart}

\usepackage{hyperref}
\usepackage{array}
\usepackage{soul} 
\usepackage{xcolor} 
\definecolor{lightblue}{RGB}{219, 236, 255} 
\sethlcolor{lightblue}

\usepackage{hyperref}
\usepackage{array}
\AtBeginDocument{%
  \providecommand\BibTeX{{%
    \normalfont B\kern-0.5em{\scshape i\kern-0.25em b}\kern-0.8em\TeX}}}

\acmISBN{978-1-4503-XXXX-X/18/06}

\begin{document}

\title[How Do Community Advocates Envision Data Intermediaries?]{“Near Data” and “Far Data” for Urban Sustainability: How Do Community Advocates Envision Data Intermediaries?}

\author{Han Qiao}
\email{h.qiao@mail.utoronto.ca}
\orcid{}
\affiliation{%
  \institution{University of Toronto}
  \streetaddress{}
  \city{Toronto}
  \state{Ontario}
  \country{Canada}
  \postcode{}
}

\author{Siyi Wu}
\email{reyna.wu@mail.utoronto.ca}
\affiliation{%
  \institution{University of Toronto}
  \streetaddress{}
  \city{Toronto}
  \state{Ontario}
  \country{Canada}
  \postcode{}
}

\author{Christoph Becker}
\email{christoph.becker@utoronto.ca}
\affiliation{%
  \institution{University of Toronto}
  \streetaddress{}
  \city{Toronto}
  \state{Ontario}
  \country{Canada}
  \postcode{}
}

\renewcommand{\shortauthors}{Qiao, et al.}

\begin{abstract}
{In the densifying data ecosystem of today’s cities, data intermediaries are crucial stakeholders in facilitating data access and use. Community advocates live in these sites of social injustices and opportunities for change. Highly experienced in working with data to enact change, they offer distinctive insights on data practices and tools. This paper examines the unique perspectives that community advocates offer on data intermediaries. Based on interviews with 17 advocates working with 23 grassroots and nonprofit organizations, we propose the quality of "near" and "far" to be seriously considered in data intermediaries' works and articulate advocates’ vision of connecting ``near data’’ and ``far data.’’ To pursue this vision, we identified three pathways for data intermediaries: align data exploration with ways of storytelling, communicate context and uncertainties, and decenter artifacts for relationship building. These pathways help data intermediaries to put data feminism into practice, surface design opportunities and tensions, and raise key questions for supporting the pursuit of the Right to the City.}
\end{abstract}





\received{2 July 2024}

\maketitle

\section{Introduction}
Cities have long been sites of social injustice and opportunities for radical changes \cite{Fainstein2014}. Many urban areas are experiencing two converging forces: rapid urbanization and datafication \cite{Currie2022}. Private companies gather sales and cellular data to predict consumer behavior \cite{green_smart_2019}, governments establish open data programs to promote transparency and encourage civic engagement \cite{matheus2014, kassen2013, doctor2022}, and grassroot community efforts organize to collect and use data to address issues of shared concern \cite{erete2016, Boone2023}. 

Urban data ecosystems encompass diverse stakeholders, data sources, various data practices, infrastructures, policies, and social and cultural factors that shape these data practices \cite{kitchin_fragmented_2021}. Among the various components forming this urban data ecosystem, \textbf{data intermediaries} -- individuals or groups providing services and tools to support others to understand and use data -- play a key role \cite{gebka2021}. Almost all stakeholders in this urban data ecosystem rely on some sort of data intermediary. Many also act as data intermediaries for others, either by sharing data in structured forms or by providing access through other means such as data visualizations, and sometimes both. As critical geographer Rob Kitchin \cite{kitchin_knowing_2015} argues, urban data does not only describe urban systems, but also shapes how a city should be governed as well as what a city should become in the future. However, whose views are represented by these data, who is empowered by the datafication process, and who has a say? The future of cities is unevenly distributed \cite{Bopp2017, greene_promise_2021}. The work of data intermediaries is central to this process and should be critically examined. 

Nonprofit and grassroots organizations are major forces of advocating for marginalized voices in urban areas and enacting change. Even though members of these groups have to navigate through disempowerment from datafication \cite{Bopp2017, Voida2011}, they overcome all kinds of hurdles so that they can continuously develop data practices that center their values and commitments. For instance, advocates focus on the affective and narrative nature of data through agonistic data practices \cite{CrooksCurrie2021}, and they prioritize values of care through data tinkering \cite{Tran2022, Boone2023}. With these unique perspectives, community advocates are experts in working with and as data intermediaries to enact change.

As Darian et al. \cite{Darian2023} argue, today’s advocacy work is, in many regards, data work. D'Ignazio \cite{D'Ignazio2022} describes community advocates as “real-world people who are using data science in the service of real-world struggles for justice”. Community advocates then have much to offer those of us researching design methods and theories for data practices and exploration tools. They show us how data intermediaries can design data exploration tools that empower marginalized voices. Anchoring on the concept of “the Right to the City” \cite{lefebvre1968}, where every inhabitant has the right to participate and shape urban futures, our work is guided by the question \textbf{what are community advocates’ perspectives, insights and visions for data intermediaries and data tools built by data intermediaries?}

In this paper, we examine the perspectives of community advocates working with grassroots and nonprofit organizations through semi-structured interviews that incorporate a technology probe. We interviewed 17 advocates working with 23 grassroots and nonprofit organizations focusing on different areas of urban sustainability in Toronto, Canada. During the interviews, we incorporated \textit{Curbcut Toronto}, a tool for exploring and visualizing urban sustainability data in Toronto designed by a group of data intermediaries, as a technology probe. Through these conversations, we learned how advocates work with a wide spectrum of data, ranging from their own lived experience to large scale open data. We recognize the distance between advocates and the data they use is important to their work, and we propose the terms \textbf{“near data”} and \textbf{“far data”} to distinguish the two extremes to help people think through this important aspect of urban data seriously considered by community advocates. Near data refers to data that is close to the everyday lives of the community advocates, including self-collected data, lived experiences, and conversations. Far data refers to heavily processed data that are often passed through many hands from other locations in the urban data ecosystem, such as open data from government or academic institutions. Through understanding advocates’ ways of using data, challenges of using data and their reflections around their experience interacting with \textit{Curbcut Toronto}, we identified one overarching design vision for data intermediaries, \textbf{connecting near data and far data}, as well as three pathways to work towards this vision: align data exploration with ways of storytelling, communicate context and uncertainties, and decenter artifacts for relationship building. Based on the vision and the pathways, we examine how advocates' visions of data intermediaries reflect commitments to data feminism. Finally, we discuss tensions that arise when connecting near data and far data, data intermediaries' roles in pursuing “the Right to the City”, and how the concept of near and far data should be emphasized as part of a dynamic yet important context for working with data. Through this study, we contribute to the growing literature on data practices in community advocacy work and develop design considerations for HCI communities and civic tech groups who research with and work as data intermediaries in support for more democratic and inclusive urban data ecosystems and urban futures.

\section{Related Work}

The foundation of our work builds on the concept of ``the Right to the City,'' which emphasizes that the right to re-imagine future urban spaces should belong to its residents and highlights  the importance of collective action and advocacy. It has found new relevance and interpretations as cities going through rapid datafication. We discuss how our work is grounded in previous work on understanding data practices in grassroots and nonprofit organizations and on exploring roles of data intermediaries in the larger data ecosystem. Our work aims to speak to both bodies of literature in the HCI community, connecting them to open up spaces for new discussions.

\subsection{“The Right to the City” Today}

The concept of “the Right to the City” (RTTC) emerged in the 1960s, when many parts of the world were going through intensified urbanization. In Lefebvre’s famous essay “the Right to the City,” he highlighted the collective struggle of urban inhabitants under accelerated urbanization pushed forward by industrial capitalism. He argued for a collective re-imagination of urban life that embodies democratization and recenters social relations \cite{lefebvre1968}. For Lefebvre, urban space goes beyond the material space to include social relationships and lived experiences of everyday life. The right to such space goes beyond citizen rights defined by the state.  It encompasses a demand for and commitment to change stemming out of collective struggles \cite{Purcell2014, lefebvre1968, harvey2013}. RTTC has a spatial highlight in cities and urban areas, sites that carry political, social and cultural significance -- the city is a site for developing capitalism but also a site for the destruction of capitalism; it is a site of struggle and conflict but also the most common site where alternative visions for our societies have emerged.

As urban areas experience rapid datafication in recent years, many scholars have been expanding and examining the meaning of  RTTC in newer contexts \cite{Anastasiu2019}. A growing body of literature examines how urban data and digital tools, like urban dashboards, prioritize reductionist and objectivist ways of knowing and governing cities that impact collective efforts towards RTTC \cite{kitchin2014, mattern2017}. An inseparable topic is recognizing the socio-digital inequalities posed by the digitization and datafication process. Past research has examined the adverse impact of datafication in urban spaces, such as security of personal data, gaps in digital literacy among different social groups, as well as lack of equal opportunities and tools to influence important matters \cite{hatuka2020}. Socio-economic factors such as educational inequality, age, gender, ethnicity and residence all contribute to socio-digital inequalities brought by digitization and datafication in cities \cite{ylipulli2023}. When the idea of data-driven urban governance becomes widespread, resource-constrained mission-driven organizations, such as nonprofits and social enterprises, are disempowered as they invest time and sacrifice expertise to respond to the pressure of becoming data-driven at the expense of their mission \cite{Bopp2017}.

Beyond presenting a foundation for us to understand the struggles brought by datafication in urban areas, RTTC also continues to serve as a call to build solidarity in the face of difference and to imagine urban futures that center democracy, diversity, justice, and sustainability. Anastasiu \cite{Anastasiu2019} elaborates RTTC in the context of the smart or digital city to argue that this right goes beyond access to data or technology, but should include “the right to both produce, manage and own all of these as part of an act of political and economic empowerment that is geared primarily towards the collective benefit and the strengthening of social relations.” Heitlinger et al. \cite{heitlinger2019} further explored how design and co-design can participate in this call for RTTC through focusing on the commons, care and biocultural diversity. Emerging work around data justice sought to foreground the political nature of datafication, challenge existing power-structure manifested in data and empower the impacted communities to frame problems and address issues related to datafication \cite{han_psychopolitics_2017, eubanks_automating_2018, dencik_conceptual_2018, oneil_weapons_2016, taylor_what_2017, noble_algorithms_2018}. Currie et al. \cite{Currie2022}, bridging the concept of data justice and RTTC in the age of ‘Big Data’, algorithms, artificial intelligence and other data-intensive technologies, highlighted the central role of people, lived experiences, and bottom-up efforts in leading radical changes. 

Shaw and Graham \cite{shaw_informational_2017} argue that data as representation of cities are “often as important as their bricks and mortar" (p908) and that those who have more control over data have more power over the city. Thus data intermediaries act as an important stakeholder in the urban data ecosystem affecting the right to access and use data, to form narratives through data, to argue for what are important urban issues, and to shape possible urban futures. HCI researchers are especially well-positioned to contribute to this intersection between design, computing, and social and political implications in the urban context.

\subsection{Data intermediaries in HCI}
Data intermediaries are individuals or groups providing services and tools for supporting others to understand and use data \cite{gebka2021}. Given the limitations and challenges surrounding the access and effective use of data, particularly for marginalized communities, data intermediaries become a critical stakeholder in this space. As Shaharudin et al. \cite{Shaharudin2023} discussed, the data intermediary role can be played by a variety of actors such as civil society bodies, journalists, entrepreneurs and researchers \cite{sawicki1996, magalhaes2013} making data more useful in various ways such as compiling data from various sources \cite{Shaharudin2023}, breaking down technical barriers \cite{Dove2023}, curating datasets based on smaller geographies such as neighborhoods \cite{yoon2018}, or offering consultation and partnership services to empower local communities in data utilization \cite{vlachokyriakos2016}. Data intermediaries serve as vital facilitators, bridging the divide between data availability and its impactful utilization \cite{Dove2023}. 

The HCI community has explored various topics relevant to the role of data intermediaries and designs of data intermediary tools. One key research area for data intermediaries is access and use of open data. Past researchers have examined what it means for open data to be accessible, developing evaluation metrics and identifying challenges and barriers for effective use \cite{frank_user_2016, Gurstein_2011, Dove2023}. They have also suggested the important role of data intermediaries in facilitating meaningful engagement with open data. Corbett et al. \cite{Corbett2018} studied open data programs in the city of Edmonton, Canada and highlighted that data intermediaries are key to the success of such efforts. Dove et al. \cite{Dove2023}, through interviewing data intermediaries,  identified pathways to successfully nurturing a community of data intermediaries. In the context of community advocacy works, other researchers highlighted the pivotal role of data intermediaries as they connect various stakeholders and bridge gaps in existing data \cite{Shaharudin2023, Puussaar2018}. 

The HCI community has also explored the design of data intermediary tools. For example, Data:In Place was an open-source platform co-designed with communities and charities to simplify access to open data for civic advocacy \cite{Puussaar2018}; Governor was an open-source web app that streamlines searching and integrating data tables in Open Government Data Platforms, with provenance tracking validated through user studies \cite{Liu2023}. Data intermediaries’ work is never just restricted to research settings. Close to the location of our study, the City of Toronto has many data intermediary projects led by local residents: OpenWaterData gathers various sources of water quality information and presents it through a web application for informing local open water swimmers and paddle-boarders \cite{openwater}, BikeSpace is a crowdsourced data platform where cyclists anonymously report bicycle parking issues across Toronto \cite{bikespace}, and No More Noise Toronto combines crowdsourced urban noise data with the city’s open information to raise awareness of urban noise pollution and advocate for policy change \cite{no_more_noise_toronto_no_2024}.

The work of data intermediaries might also carry risks such as deteriorating data quality, reinforcing bias, oppressing marginalized voices, and thus leading to unethical or harmful uses of data. For instance, as data is processed by numerous layers of data intermediaries, White \cite{white_following_2017} illustrated how data can be used and interpreted in various ways not anticipated with respect to their original generation. Bates et al. \cite{bates_data_2016} proposed the concept of data journey to emphasize the importance of recognizing the physical sites and socio-material conditions of data practices as layers of data intermediaries pass down the data. When analyzing data practices through the lens of data journeys, Jensen et al. \cite{jensen_tensions_2023} also revealed challenges, such as assessing data validity through standardized data algorithms versus competencies shared socially. While the work done by data intermediaries can bridge access and use of various data, they might not necessarily be inclusive of local community members \cite{choi2017}.

Therefore data intermediaries play key roles in making data better or worse, and their work should be examined carefully. CSCW and HCI researchers exploring design methods and data practices have called for collective sense-making and co-design in data work \cite{seidelin_co-designing_2020, neff_critique_2017}. Community-oriented citizen science researchers also emphasized the importance of co-design with local community, highlighting crucial values in democratic participation, local knowledge, and equitable power relationships in enacting social change \cite{hsu_smell_2019, chari_promise_2017, paulos_citizen_2008, Preece_2019, Hsu_2018}. Therefore, community advocates’ perspectives on data intermediaries hold much value as they represent and amplify lived experiences and marginalized voices.

\subsection{Data Practices in Grassroots and Nonprofit Organizations}
Many scholars have examined inequalities caused by datafication \cite{hatuka2020, ylipulli2023} and challenges experienced by grassroots and nonprofit organizations \cite{Bopp2017}. A growing body of research in HCI also highlighted the creative and attentive ways of using data when these groups navigate through the constraints and tensions posed by datafication \cite{Boone2023, Tran2022, D'Ignazio2022}. As D’Ignazio \citep[p.10]{D'Ignazio2022} wrote: “grassroots data activists at the margins -- real-world people who are using data science in the service of real-world struggles for justice -- have much more expertise to offer those of us trying to craft data frameworks and practices in the service of justice.”

Grassroots and nonprofit organizations are leveraging data in diverse ways. These organizations use open data for storytelling to achieve goals such as grant applications, internal evaluations, and obtaining support from local government officials \cite{erete2016}. Community advocacy groups leverage data for legitimizing the organization’s expertise and mission, mobilizing audiences, promoting innovative work, and amplifying the visibility of marginalized beneficiaries \cite{Darian2023}. Under the overarching goal of legitimization, community organizers leverage data in diverse forms of legitimation towards different audiences, such as “legitimizing the need for change within the community,” “legitimizing claims against the state,” and “legitimizing lived experience” \cite{pei2022}.

Grassroots and nonprofit organizations are exemplary role models for pushing against the norms of a data economy conventionally organized around science, surveillance and selling \cite{dignazioklein2021}. These organizations, on the other hand, organize around social good and goals of co-liberation that strive for the values, goals and voices of their beneficiaries \cite{dignazioklein2021, Darian2023}. Their work is also usually an enactment of care, a focus beyond productivity to prioritization of relationships, social and cultural community context, and attentiveness to ways of nurturing and sustaining one another \cite{Tran2022, Boone2023}.

Researchers also highlighted how grassroots and nonprofit organizations navigated the challenge of double bind, contradictory situations and tensions when using data \cite{CrooksCurrie2021, fortun2001}. For example, feminist activists adopt a practice of “refusing while using data and using while refusing data” when documenting stories related to feminicides as they navigate tensions carried by quantitative data science \cite{D'Ignazio2022}. Community organizers also navigated the tension between protecting their communities from surveillance through data and securing funds through reporting data to funders \cite{pei2022}. Crooks and Currie \cite{CrooksCurrie2021} proposed the concept of agonistic data practices, which recognizes the limitations of using quantification and offers a liberating framework that centers affective and narrative ways of using data. Darian et al.\cite{Darian2023} showed how advocates enact the seven principles of data feminism \cite{dignazioklein2021}, listed in Table \ref{tab:data_feminism}, which we will refer to in our analysis.

\begin{table}[h!]
\centering
\begin{tabular}{p{0.4\linewidth}|p{0.5\linewidth}}
\hline
\textbf{Data Feminism Principle} & \textbf{This principle means...} \\
\hline
Examine Power (DF1) & Recognize forces of oppression shaping urban lives and reflected in urban data and data tools. \\
\hline
Challenge Power (DF2) & Push back against unequal power structures to work towards alternative urban futures. \\
\hline
Elevate Emotion and Embodiment (DF3) & Recognize that knowledge about urban spaces comes from inhabitants' lived and bodily experiences. \\
\hline
Rethink Binaries and Hierarchies (DF4) & Challenge systems of counting and classification often found in data describing cities. \\
\hline
Embrace Pluralism (DF5) & Value multiple perspectives and diverse ways of knowing a city and using urban data. \\
\hline
Consider Context (DF6) & Attend to the provenance and environment from which the data was collected and processed, recognizing that data are never objective representations of urban lives. \\
\hline
Make Labor Visible (DF7) & Value and make visible the work of those who collect, process, annotate, and curate urban data. \\
\hline
\end{tabular}
\caption{Data feminism principles and their explanations in the context of urban data.}
\label{tab:data_feminism}
\end{table}

Ultimately, data advocacy work goes far beyond data. A successful campaign or mobilization requires social and political capital \cite{D'Ignazio2022, CrooksCurrie2021}. As grassroots and nonprofit organizations fight in the frontline of social injustice and adopt powerful ways of leveraging data in challenging situations, they form a unique lens towards data for advocacy that has much to offer those of us working in the HCI research community, hoping to understand design and democratization of data support tools and practices. Thus, we talk to and learn from community advocates working with grassroots and nonprofit organizations to understand their insights on working with and as data intermediaries.

\section{Method}
\subsection{Participants and Sampling Method}

We interviewed 17 participants working with 23 grassroots and nonprofit organizations focusing on urban sustainability issues in Toronto neighborhoods. Their focal topics range from affordable housing, transit access, road safety, mental health, climate, park and food access etc. For instance, P1's focus on urban noise addresses an urban environmental health hazard that impacts residents' quality of life, and P5's work on youth engagement in urban planning promotes inclusivity and long-term community resilience, both essential to urban sustainability. We reference all participants by number P1 to P17 and document detailed information about their topics of work in Table \ref{table1}. 

Participants were initially recruited from the first author’s existing network stemming out of participation in a local civic tech organization for two years, then through snowball sampling as well as cold emailing through organizations’ websites. Contact information was collected through a scan of Toronto’s urban sustainability focused community groups, including environment, housing, transportation and health, which are not only key issues in urban sustainability, but also the four data topics included in the current version of \textit{Curbcut Toronto}, the technology probe we included in the interviews. When recruiting participants, we specifically focused on recruiting participants who think data already influences their approach to community advocacy work.

Among the participants, eight identified as female, two identified as non-binary and seven identified as male. Ten of the 17 participants are involved in more than one community group focusing on different yet interrelated topics around urban sustainability. P7 explained that all the different kinds of community advocacy work \textit{“synergize”} with each other and then elaborated that \textit{“many advocates are wearing multiple hats with multiple lenses, but all working towards hopefully the same goal of creating safer spaces and safer means to get to those spaces and places throughout the city”} and that \textit{“they all work together in a way that looks at it with an equitable lens.”} To categorize the type of advocacy work they do, we adopted Frumkin’s framework \cite{frumkin2005}, which was also used in \cite{Darian2023}: seven of the participants have a focus on civic and political engagement, five have a focus on service delivery and the other five cover both types of work. The participants self-identified the type of organization that they work with as either more informal community organizing, which we refer to as grassroots organizations, or nonprofit organizations. Previous works in HCI have discussed both types of organizations together \cite{Tran2022} and focused on each type separately \cite{pei2022, erete2016}. In our project, both types of organizations share more similarities than differences when working with data for advocacy. Among our participants, all are users of data intermediaries' work and some act as data intermediaries. Initially, we recruited participants with a focus on community advocates as potential users of data tools developed by data intermediaries. However, through our interviews, we discovered that many participants not only work with data provided by data intermediaries but also contribute to the work of data intermediaries in various ways. For example, P3 uses data from intermediaries and combines it with their own data for storytelling through local magazines, while P1, in addition to using works done by data intermediaries, also acts as a data intermediary by collecting counter-data \cite{D'Ignazio2022} and providing data visualizations to the broader public.

\begin{table}[h]
    \centering
    \begin{tabular}{p{0.1\textwidth}>{\raggedright}p{0.2\textwidth}
    p{0.25\textwidth}>{\raggedright\arraybackslash}p{0.25\textwidth}}
        \hline
        Participant & Advocacy topic & Type of advocacy work & Type of organization \\
        \hline
        P1  & Urban noise, biking and bike safety & Civic \& Political Engagement & Grassroots organization \\
        \hline
        P2  & Road safety, park access & Civic \& Political Engagement & Grassroots organization \\
        \hline
        P3  & Mental health, affordable housing, public transit & Civic \& Political Engagement, Service Delivery & Grassroots organization, Nonprofit organization \\
        \hline
        P4  & Climate & Civic \& Political Engagement, Service Delivery & Grassroots organization \\
        \hline
        P5  & Road safety, youth engagement in urban planning & Civic \& Political Engagement & Grassroots organization \\
        \hline
        P6  & Public space activation & Service Delivery & Nonprofit organization \\
        \hline
        P7  & Bike safety, library, public space activation & Civic \& Political Engagement, Service Delivery & Grassroots organization, Nonprofit organization \\
        \hline
        P8  & Public transit, accessibility & Civic \& Political Engagement & Grassroots organization \\
        \hline
        P9  & Park access, environment & Service Delivery & Nonprofit organization \\
        \hline
        P10 & Youth engagement in urban planning & Service Delivery & Nonprofit organization \\
        \hline
        P11 & Public health, civic engagement & Civic \& Political Engagement & Grassroots organization \\
        \hline
        P12 & Park access, food access & Civic \& Political Engagement, Service Delivery & Grassroots organization, Nonprofit organization \\
        \hline
        P13 & Employment opportunity access & Service Delivery & Nonprofit organization \\
        \hline
        P14 & Shelter services & Service Delivery & Nonprofit organization \\
        \hline
        P15 & Cultural resilience & Service Delivery & Nonprofit organization \\
        \hline
        P16 & Climate & Civic \& Political Engagement & Grassroots organization \\
        \hline
        P17 & Climate, food sovereignty & Civic \& Political Engagement, Service Delivery & Nonprofit organization \\
        \hline
    \end{tabular}
    \caption{Summary of Participants and Their Advocacy Work}
    \label{table1}
\end{table}

\subsection{Data Collection}

We collected data through a series of semi-structured interviews remotely through a video conferencing platform, Zoom. All interviews were conducted by the first author, recorded and transcribed by both the first and the second author for further analysis. The interviews lasted between 28 and 96 minutes, accumulating a total duration of 16 hours and 29 minutes, with a median length of 53 minutes and an average of 58 minutes. All interviews were conducted between April and November 2023. Two of the interviews were conducted in two sittings within a week due to participants’ time availability and eagerness to chat more than the expected time.

The interviews were conducted based on a protocol that consisted of three main parts:
1) the community advocacy work that the participant is involved in 2) their experiences of using or not using data in their work 3) interacting with \textit{Curbcut Toronto}, a data exploration and visualization tool, and sharing their thoughts and reflections around using data intermediary tools. During each interview, follow-up questions were asked based on participants’ specific experiences.

For the third part, when \textit{Curbcut Toronto} came in as a technology probe, participants were asked to share their screen through the video conferencing tool while interacting with the tool. Before sharing the tool to participants, the interviewer explained the think-aloud method \cite{alhadreti2018, gill2012} and then asked the participants to freely explore the tool while thinking aloud. After the participants felt that they had fully explored the tool, the interviewer continued the semi-structured interview, asking follow up questions based on the interview protocol to prompt users to reflect on their experiences using the tool. 

\subsection{\textit{Curbcut Toronto} as a Data Intermediary Technology Probe}
To better understand community advocates’ insights and perspectives around data intermediaries, we incorporated \textit{Curbcut Toronto} as a technology probe during our interviews to provoke thoughts and help participants articulate their insights around data intermediary tools. 

\textit{Curbcut Toronto} is an open-source urban data exploration and visualization tool, aiming to support urban data access and use. The original version of \textit{Curbcut} was developed for Montreal, Canada by the McGill University Sustainability Systems Initiative, aiming to support flexible and comprehensive exploration of urban sustainability data. A Toronto version of the tool was launched in partnership with the University of Toronto Digital Curation Institute.

\begin{figure*}[h]
  \centering
  \includegraphics[width=\textwidth]{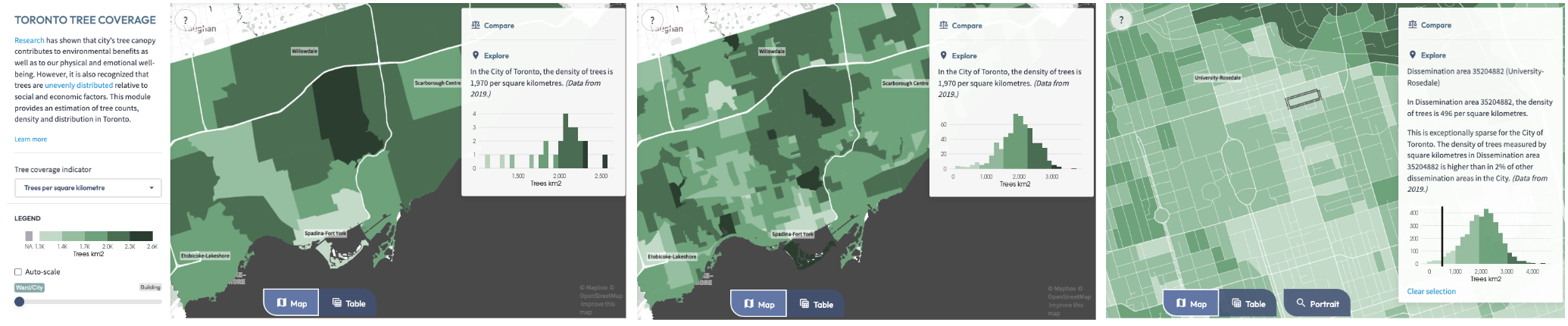}
  \caption{Screenshots of \textit{Curbcut Toronto} visualizing tree coverage in Toronto in three different geographic scales. Left: an example of visualization in ward level. Middle: an example of visualization in census tract level. Right: an example of visualization in dissemination area level.} 
  \Description{Three screenshots of the data exploration and visualization tool Curbcut Toronto presented side by side. On the Left, there is a screenshot showing a visualization of tree coverage in Toronto for each ward. The screenshot contains a choropleth map in different shades of green where the darker the green, the higher number of trees per square kilometer. In the middle, there is a screenshot showing the same information but zooming into census tract level, where the blocks are granular. Lastly on the right, there is a screenshot zooming into the scale of dissemination areas, which is the smallest geographic scale that the system supports.}
  \label{fig:teaser}
\end{figure*}

\begin{figure*}[h]
  \centering
  \includegraphics[width=\textwidth]{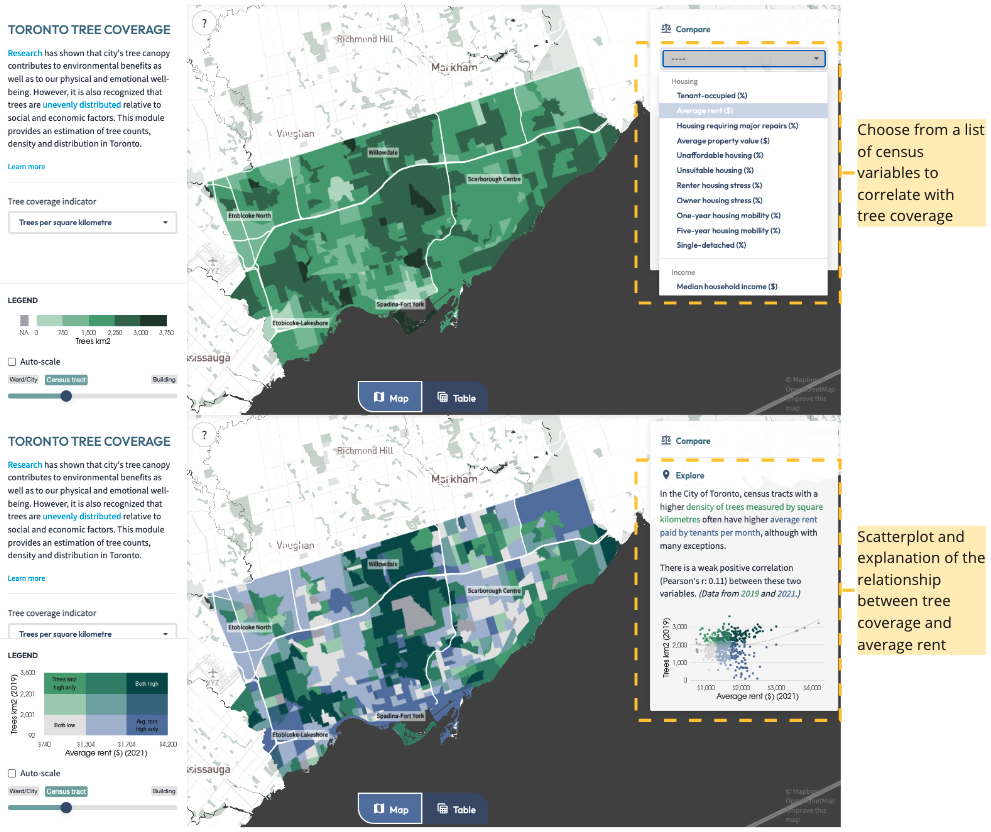}
  \caption
  {\textit{Curbcut Toronto} allows users to correlate the featured variable on the map with a census variable. In the top screenshot, the users can select a variable from a list of census variables. In the bottom screenshot, the user is visualizing the correlation between Toronto tree coverage and average rent. A description text box is showing on the right indicating that there is a weak positive correlation between the two variables.}
  \Description{Two screenshots of Curbcut Toronto. On the top, one screenshot shows a visualization of tree coverage in Toronto for each census tract. The screenshot contains a choropleth map in different shades of green where the darker the green, the higher number of trees per square kilometer. On the left side of the visualization, a drop down menu is clicked open presenting a list of census variables, including tenant-occupied percentage, average rent, average property value and so on. A note is written on the side indicating that a user can choose a variable from this list of census variables to visualize its correlation with tree coverage. On the bottom, another screenshot shows a visualization of the correlation between Toronto tree coverage and average rent. The visualization adopts two color schemes: green indicates tree coverage and cyan indicates average rent. On the right side of the visualization, another notes point to a scatter plot and explanation of the relationship between tree coverage and average rent. The description says: In the City of Toronto, census tracts with a higher density of trees measured by square kilometres often have higher average rent paid by tenants per month, although with many exceptions. There is a weak positive correlation. Pearson's r: 0.11. between these two variables. Data from 2019 and 2021.}
  \label{fig:curbcut_correlate}
\end{figure*}

\begin{figure*}[h]
  \centering
  \includegraphics[width=\textwidth]{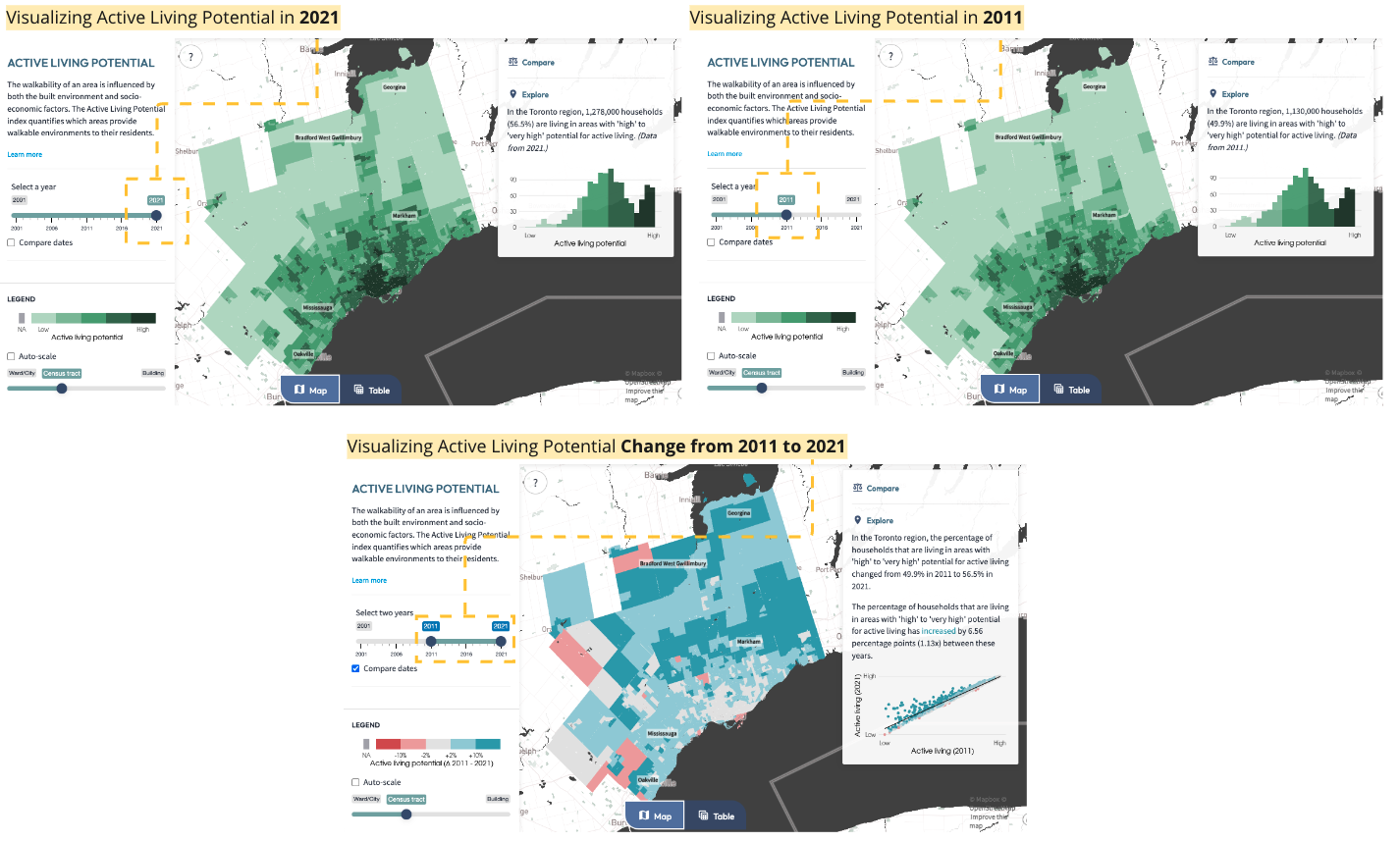}
  \caption
  {Temporal exploration features of \textit{Curbcut Toronto}. The tool allows users to look through available historic data and compare data at two different dates. Top left: visualizing active living potential in 2021. Top right: visualizing active living potential in 2011. Bottom: visualizing active living potential change from 2011 to 2021.}
  
  \Description{Three screenshots of Curbcut Toronto. On the top, two screenshots are placed side by side, visualizing active living potential data in the Greater Toronto Area in two different years. The left screenshot is visualizing the data from 2021 and the right screenshot is visualizing the data from 2011. Both visualizations are choropleth maps with different shades of green as its color. The darker the green, the higher the active living potential in a census tract. On the bottom, one screenshot is visualizing the change of active living potential between 2011 and 2021. The visualization is a choropleth map with different shades of blue and red, where blue indicates an increase in active living potential and read indicates a decrease in active living potential.}
  \label{fig:curbcut_temporal}
\end{figure*}

\textit{Curbcut Toronto} offers map-based visualizations for data related to various urban sustainability topics, including environment, health, housing, and transportation. All data comes from open data sources including City of Toronto’s Open Data Portal \cite{city_of_toronto_toronto_2017}, Canadian census data \cite{government_of_canada_census_2001}, Canadian Mortgage and Housing Corporation \cite{cmhc_canada_nodate}, the Canadian Urban Environmental Health Research Consortium \cite{canue_canadian_2023} as well as OpenStreetMap \cite{openstreetmap}. The tool offers visualizations for one variable in different geographical scales as shown in Fig \ref{fig:teaser}, bivariable relationships between two variables as shown in Fig \ref{fig:curbcut_correlate}, and visualizations for temporal changes as shown in Fig \ref{fig:curbcut_temporal}. \textit{Curbcut Toronto} represents in many ways a common type of data intermediary tool: it combines open data from various sources in new ways, making them more accessible through visualizations and explanatory texts. It  differs from other map-based visualizations tools by providing more flexible ways of visualization and analysis.

\begin{figure*}[h]
  \centering
  \includegraphics[width=0.8\textwidth]{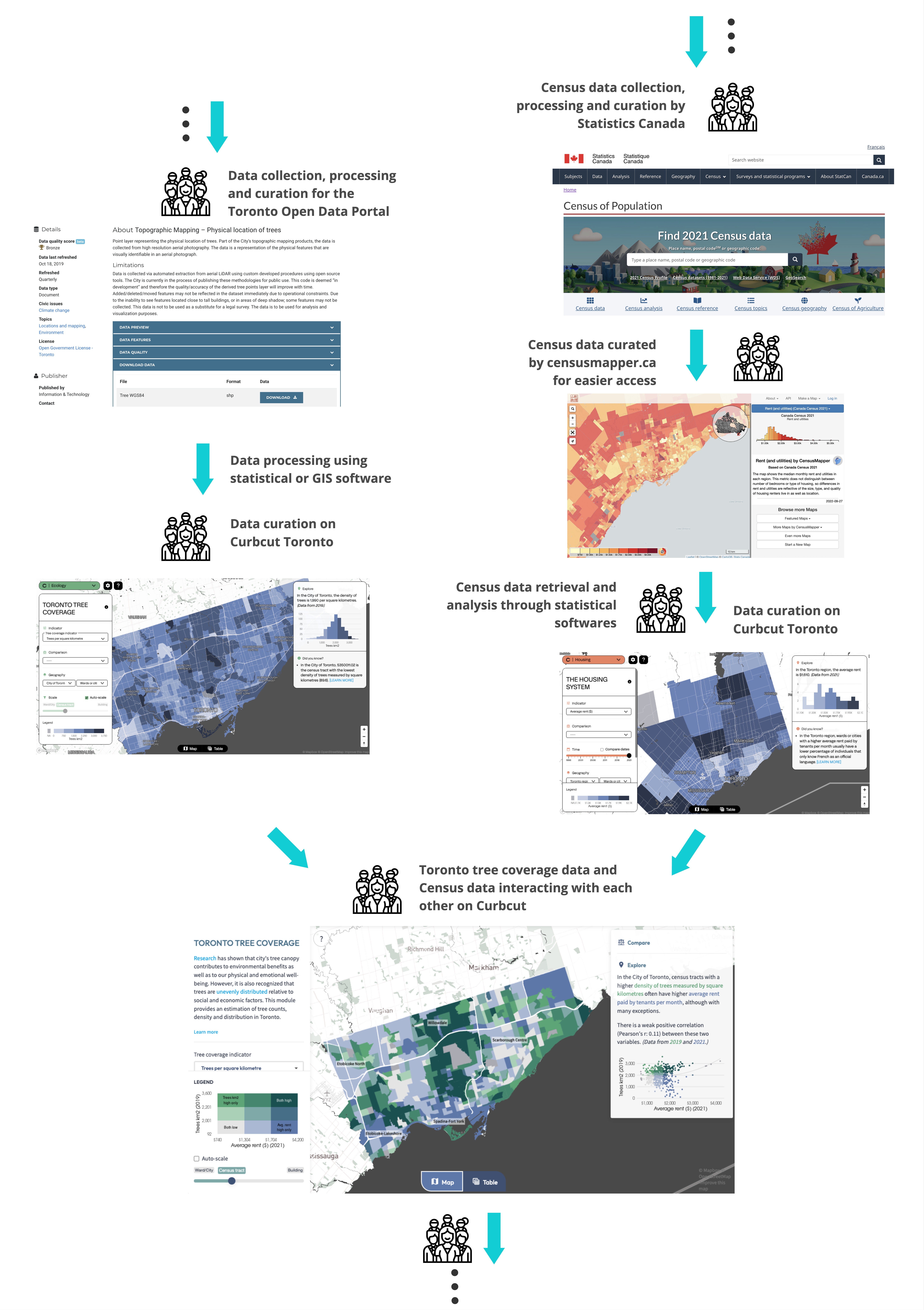}
  \caption
  {An illustration of a segment within the urban data ecosystem focusing on how data passes through data intermediaries to get to \textit{Curbcut Toronto}. Data intermediaries processing and curating data before it is presented on \textit{Curbcut Toronto} include people working on the Toronto Open Data Portal, Statistics Canada, CensusMapper and so on. Designers of \textit{Curbcut Toronto} act as another data intermediary. Data from Curbcut Toronto can then passed on to other data workers for future uses.}
  \Description{Multiple screenshots of data intermediary websites, including Curbcut Toronto and Census Canada are displayed in two columns. The screenshots illustrate how data goes through layers of data intermediaries to get to Curbcut Toronto. The screenshots are being organized into two main columns, connected by cyan colored arrows and human icons in between. Starting from the first screenshot of the first column on the left side, a cyan arrow points to three dots representing the multiple layers of data intermediaries that are not captured in this figure. The first screenshot on the top left is a screenshot from Toronto Open Data Portal with an annotation on its top right saying data collection, processing and curation for the Toronto Open Data Portal. Then a cyan arrow and a human icon leads to the second screenshot of the Toronto tree coverage module from Curbcut Toronto. On the top right corner of this screenshots is an annotation saying data processing using statistical or GIS software and data curation on Curbcut Toronto. In parallel with the first column, the second column on the right side starts with a cyan arrow, three dots and a human icon pointing to a screenshot from Statistics Canada. On the top left side of the screenshot, an annotation says census data collection, processing and curation by Statistics Canada. Below that shows a screenshot from censusmapper.ca with an annotation on the top left side says census data curated by censusmapper.ca for easier access. Below, is a human icon with an annotation that says census data retrieval and analysis through statistical software. Left to this annotation is another cyan arrow that points to the second screenshot of Curbcut Toronto in the first column. The arrow is annotated as data curation on Curbcut Toronto and this illustrates how data from various sources are now being brought together onto Curbcut Toronto. Below this is another cyan arrow and a human icon pointing to the last screenshot of this figure. This screenshot of Curbcut Toronto illustrates Toronto tree coverage data and census data interacting with each other on Curbcut through a visualization. The screenshots shows a visualization of the correlation between Toronto tree coverage and average rent. Below this screenshot is another human icon, a cyan arrow and three dots, illustrating that these data on Curbcut will then be passed down for future uses.}
  \label{fig:curbcut_data_journey}
\end{figure*}

Like other systems of this kind \cite{manhattan, healthyplan,electricity_2017}, \textit{Curbcut Toronto} relies on layers of data intermediaries curating, preparing and processing data from a wide range of sources (Fig \ref{fig:curbcut_data_journey}). For instance, the visualization module on Toronto tree coverage is prepared through incorporating the Canadian census data \cite{government_of_canada_census_2001} and Topographic Mapping of Trees data from Toronto Open Data Portal \cite{city_of_toronto_toronto_2017}. For these steps to be possible,  many layers of data intermediaries already worked to collect, process and curate these data. The data presented on \textit{Curbcut} will also be passed on to future layers of data intermediaries through the export features that allow users to reuse the platform's datasets.

We chose to incorporate \textit{Curbcut Toronto} in our semi-structured interviews as a technology probe because it offers a lens for provoking reflections and insights. The HCI and design community have had rich experiences adopting variations of cultural probes since its introduction by Gaver et al \cite{gaver1999}. Since then, technology probe has become one of the most popular variations of the original method \cite{boehner2007}, where a technology artifact is used to collect information about the use and user of a technology and inspire new designs to support users’ needs \cite{hutchinson2003}. Boehner et al. \cite{boehner2007} summarized that, within the HCI community, technology probes are used for various purposes from collecting information around usability for a single design to providing inspiration for opening up new design spaces. In our study, \textit{Curbcut Toronto} serves more towards the latter purpose, as we hope to examine the general design space and visions for data intermediary tools as opposed to formulating a rigid design guideline for data intermediary tools to fulfill. Echoing previous works \cite{jorke2023, funk2021}, we chose to use \textit{Curbcut Toronto} as a lens to provoke reflections and reactions, as well as to support participants to articulate thoughts, feedback and questions. 

\subsection{Data Analysis}

We adopted an inductive thematic analysis approach \cite{braun2006} to analyze the interview videos and transcripts, to iteratively form codes for understanding community advocacy groups’ existing ways of using data and their visions for data intermediary tools. 

The first and second author watched the interview recordings to transcribe interviews with notes and screenshots of how participants interacted with \textit{Curbcut Toronto}. For each interview, the first and second author would first watch recordings, read and re-read the transcripts to generate initial codes and then come together in open coding sessions to read through the transcripts again, watch recordings when necessary, discuss codes and finalize the initial rounds of codes that were recorded using a coding software, ATLAS.ti.

After we conducted seven interviews, we documented five emergent themes, including the nature of advocacy work, ways of using data, feelings towards data, ways of using data intermediaries, and challenges of using data intermediaries. As we continued conducting more interviews, we also returned to literature around the topics of theoretical work of understanding urban data and open government data \cite{Puussaar2018, kitchin2014, mattern2015}, data practices in community works \cite{Boone2023, Bopp2017, D'Ignazio2022}, and design practices used in data intermediary tools \cite{D'Ignazio2022, erete2016, Shaharudin2023}. We iteratively went back to refine our codes and as we conducted 11 interviews, we transferred all codes to a Miro board, an online collaborative visualization platform, as sticky notes, and conducted iterations of affinity diagramming to summarize emergent themes around community groups’ ongoing ways of working with data and their visions for data intermediaries. Informed by existing literature, emerging codes and themes from interviews, all authors came together to discuss four key insights in the result section below that illustrate the visions for data intermediaries from the perspectives of community advocates. As the relevance of data feminism principles emerged, the first author mapped each of the principles to all participants' quotes relevant to the four key themes and reviewed the mapping and codes with the last author. 

The final result was shared with participants through a paper draft with each participant's quotes and insights derived from their interview highlighted as well as a recorded video presentation. Participants were provided opportunity for comments and corrections. We received feedback like ``\textit{so nice to hear from you and read this report and watch your video (very helpful!)}'', ``\textit{It looks and sounds great. Congratulations on getting it to this point. I'm happy to have played a very small part in it}'', ``\textit{Thank you so much! I really appreciate the video and the time you took to highlight my quotes}", and no corrections were made.

\section{The Vision and the Pathways for Data Intermediaries}

From these interviews with community advocates, we identified four key themes around advocates’ perspectives for data intermediary: an overarching vision of connecting \textit{near data} and \textit{far data} and three pathways for pursuing the vision. The vision and the pathways arise from community advocates’ existing ways of working with data, tying closely to community advocates’ objectives of challenging the status quo and re-imagining alternative urban futures. In the next sections, we discuss advocates' current ways of working with data, how that connects with their vision for data intermediaries, and how the pathways for pursuing this vision support data intermediaries to put data feminism principles into practice. We labeled the emphasis on each of the seven data feminism principles throughout our result section as DF1 to DF7, corresponding to the list in related works section 2.3. 

\subsection{The Vision: Connect “near data” and “far data”} \label{sec:vision}
Community advocates shared their complex consideration of using and interacting with data they have direct experience with and know very well, as well as data obtained from distant platforms and stakeholders. We propose the concept of “near data” to refer to data that is close to community advocates' everyday lives, including self-collected data, lived experiences, and conversations. We use “far data” to refer to heavily processed data from other people, such as open data from government or academic institutions. We recognize that community advocates use a variety of data on the spectrum of near and far and that they view the distance of data as an important context for their advocacy work. Our concept of near and far builds on feminist critiques on how oftentimes data is presented as facts collected from a distance through what Haraway calls “the god trick,” when in fact it represents perspectives from dominant groups \cite{haraway1988, D'Ignazio2022}. This concept of near and far data is also constructed as a response to D’Ignazio and Klein’s data feminism call to examine and challenge power when working with data \cite{D'Ignazio2022}. While community advocates actively use both near data and far data, they treat these data differently and have different ways of working with different types of data. Advocates emphasize the importance of near data to make sense of, complement, question, and challenge far data. From their views emerges a vision for data intermediaries to facilitate engagement with both near data and far data, connecting the two to balance both, empower near, and better use all data.

\subsubsection{How advocates engage with far data}
Advocates use far data in their daily operations, for example when they create reports or write grant applications, to furnish a broader statistical context and substantiate their claims or support their arguments. Almost all participants mentioned their various ways of using census data, City of Toronto’s open data portal, and research data from academic institutions. Resonating with findings from Pei et al. \cite{pei2022}, advocates use far data to legitimize their advocacy works to a variety of audiences. For example, P9 shared using census data to provide contexts for legitimizing their service delivery work to funders: \textit{“I used Statistics Canada a lot. So in the reports I just have a general introduction to the park context, where I provide some context around the neighborhood that the parks are located in.”} Similarly, P12 mentioned their use of open data in grant applications \textit{“when our grant writer applies for funding for our project, they use open data to find information about park usage.”} P17 shared that when communicating with their community members, they would leverage research data from both the city as well as journal articles from the global South, where a lot of experts on food sovereignty are located. P6 directly acknowledges that \textit{“we rely heavily on open public data and we're very thankful for it.”} 

However, engaging with far data has presented several shared challenges among advocates. One of the most recurrent concerns is access. P1 shared their frustration, commenting that \textit{“how does somebody like me who's fighting for safe streets get access to this data? And the answer was, well, you can't.”} This frustration is also shared by P3, who finds it puzzling that after a third-party initiated environmental study on a local subway line, \textit{“they won't let us see that information.”} Financial constraints act as another substantial barrier. P4 mentioned \textit{“Esri makes it free for students. But let me tell you, it is not free, even with a nonprofit. It's really expensive.”} Many organizations are compelled to explore free alternatives, which may not offer the same quality as their paid counterparts, hindering advocates from effectively making use of open data for their work. In addition to these concerns, both P9 and P16 pointed out that scientific measurements often fail to resonate with communities and that an important role of advocates is to translate results from statistical modeling to information that their community members can easily interpret and understand. These stories collectively illustrate concepts proposed by previous researchers on minoritization, disempowerment and inequity brought by datafication and over-reliance on far data \cite{benjamin_race_2019, crooks_accesso_2019, pei2022, kitchin2014, Bopp2017}.

\subsubsection{Connect near data and far data} While advocates rely on far data, they see more value in near data to help them make sense of, enrich, question and challenge far data. Almost all participants mentioned relying on their embodied experience to understand the more abstract far data. P16 mentioned how community members made sense of the usually referenced two degrees in climate change relying on their lived experiences (DF3): \textit{“I don't understand what two degrees means. I understand it being so hot that I can't get out of my apartment and all I can do is lie down.”} Advocates also highlighted how near data enriches far data through contextualizing the abstract, making data more tangible and relatable to wider audiences (DF3, DF6). P1 argues that when near data is weaved in, \textit{“the story is not just flat,”} but extends to \textit{“how people feel, how people see, how people approach their day.”} P7 connected near data and far data through filming hours of videos of themselves and their children riding on bike lanes that are in need of improvement alongside analysis of existing traffic collisions collected by the Toronto Police Service to advocate for safer biking infrastructure and street design for marginalized road users (DF3, DF5). 

Advocates also connect near data and far data to question and challenge dominant views. Through P1’s own traffic collision accident while biking, they found out that their data is missing in the publicly available data set and started to question and challenge the far data, leading to their initiative on a \textit{counterdata} \cite{D'Ignazio2022} project called “All Road Crash Victims Count,” to allow victims of road accidents to self-report their experiences that are not captured by official sources (DF1, DF2, DF4). Similarly, P2, together with other residents in their neighborhood, identified missing speed camera data of their neighborhood in open datasets and advocated for speed reduction through referencing their years of knowledge living in the neighborhood (DF2, DF5). P11 connected their own experience navigating the healthcare system with public data on municipal political donations records to initiate projects on analyzing inequity and injustice in the public health system (DF1, DF2, DF3).

Beyond the direct usage of near data to complement or challenge far data, advocates also engage with near data through more affective and embodied ways. As P1 expressed, \textit{“it's still my data,”} emphasizing the personal and emotive attachment advocates have to their self-collected data and lived experience (DF3). They explained further, \textit{“it's very rewarding to have that relationship and then to enrich that data and be able to weave friends and that qualitative piece into it.”} P3 relies on their lived experience to tell their own stories. They remember every digit of their near data: \textit{“I received a supplement every month for 158.25 with my disability payment to cover in those days…I didn't have to think about should I spend 7.50, 3.25 one way and 3.25 to go to therapy, or am I going to eat today.”} When using near data, advocates feel their stories are more compelling, \textit{“I can talk about those numbers. I don't have to look at a chart. The charts are in my head because I lived it.”} P10 emphasized the need for more near data, as they offer depth, rooted in lived experience, but still lacking, \textit{“there are real, lived experiences and there are real people who live in these areas and whose stories are not documented.”}

With the ongoing effort in connecting near data and far data, community advocates share a vision for data intermediaries to support such endeavors. When advocates interact with \textit{Curbcut Toronto}, a tool that only provides far data, this vision of linking near data and far data was repeatedly highlighted. Many advocates discussed how data intermediaries could provide technical support for connecting near data and far data. Advocates also highlighted that data intermediaries would add much value in making advocates aware of available far data so that community advocates could better use or examine problems within these data with their near data (DF1). Furthermore, advocates emphasized the value of data intermediaries fostering local networks of community groups collecting near data to multiply existing efforts (DF7, DF1, DF2). Thus, from our conversations with community advocates, we summarize and discuss three potential pathways for data intermediaries to pursue the vision of connecting near data and far data -- (1) align data exploration with diverse ways of storytelling, (2) communicate context and uncertainties, and (3) decenter artifacts for relationship building.

\subsection{Pathway 1: Align data exploration with diverse ways of storytelling}
Storytelling in advocacy serves as a potent way to weave together data and lived experiences to create compelling narratives \cite{pei2022, erete2016}. Community advocates from our interviews shared how they leverage data to form narratives in diverse ways for different audiences. Storytelling remains a center piece for engaging civic participation and influencing social change. Advocates’ creative and thoughtful ways of using data in their storytelling also shed light on how data intermediaries could better align data exploration with diverse ways of storytelling.

\subsubsection{Support access and exploration of far data} By walking us through the diverse ways of considering and using data, community advocates offered suggestions for data intermediaries to bridge technical gaps in data access and exploration for storytelling so that far data could be brought closer to advocates’ data works. Many advocates (P5, P6, P10, P11) mentioned that when crafting their stories, they hope to know what data is available, but oftentimes they might not be aware of relevant far data that have been made available. As P11 emphasized, they hope to approach far data in an exploratory and flexible approach, not to \textit{“prove [their] theory”} but to \textit{“poke around and find a comprehensive picture”} and to be \textit{“always open to find other things in the data”} (DF1, DF5). Thus, data intermediaries can play important roles in making far data more accessible and pointing advocates to possible data related to their work.

Almost all advocates communicate their insights and stories through reports and presentations that benefit from incorporating visualizations and maps. For example, P6 talked about storytelling through a \textit{“final report in which we have a bunch of visualized data,”} where a lot of effort is put into resonating with their audiences. P7 shared that mapping brings a \textit{“more street-level understanding and a more micro understanding of the work that I do”}. By connecting abstract data to concrete contexts, mapping makes the narratives tangible and relatable for their audiences (DF3). Community advocates found \textit{Curbcut Toronto} to be useful when they could leverage visual ways of understanding the city and telling their stories. P9 shared that \textit{“a lot of nonprofits don't have people who know how to use GIS, or if they do, they don't have the licensing because it's expensive. And so having maps like this that we could use for reports would be super helpful.”} Similarly, P8 commented \textit{“[By] adding colors on a map and knowing what areas are especially affected by this issue a lot, it really helps bring in the point and leads my curiosity into knowing more topics about it.”}

Quantitative analysis support such as temporal, geographical comparison and correlation was also highlighted to be useful for supporting advocates in their storytelling. P13 commented \textit{“I think my favorite part is the ability to compare changes across time in a way that is easy.”} With a longitudinal perspective offered in data intermediary tools, advocates could easily enrich their narratives through examining how urban spaces have evolved over time, as P14 commented, \textit{“if we know, for instance, trends around homelessness, maybe one year homelessness, increased or decreased. And then you can take a look at something like what was the housing cost at that time.”} Beyond temporal comparisons, advocates (P5, P8, P14, P15) also make geographical comparisons between one neighborhood and another to add new dimensions to their storytelling. P5 noted, \textit{“so if I saw this bike comfort data, I could maybe compare it to another town.”} P12 also talked about comparing infrastructure spending in affluent neighborhoods with less privileged areas to advocate for more funding for their local neighborhoods (DF1, DF2). P2 pointed out the value of having access to various geographic scales of data, because it allows users to ask critical questions like \textit{“why does this disparity exist”} and \textit{“how can we address it”} (DF1, DF2). Offering flexibility in exploring data in different temporal and geographic scales enhances advocates' existing ways of telling local stories. Another technical support from data intermediaries could be correlations of different data. P9 was advocating for more greenspace by connecting it to health benefits in their reports: \textit{“there's this whole movement around understanding parks and green spaces’ role in social prescribing.”} P11 also leveraged correlation between household income and voting patterns to reveal underlying stories of their local provincial election. These technical ways of support align with community advocates' existing ways of storytelling. 

\subsubsection{Center marginalized ways of storytelling} 
\begin{figure*}[h]
  \centering
  \includegraphics[width=0.7\textwidth]{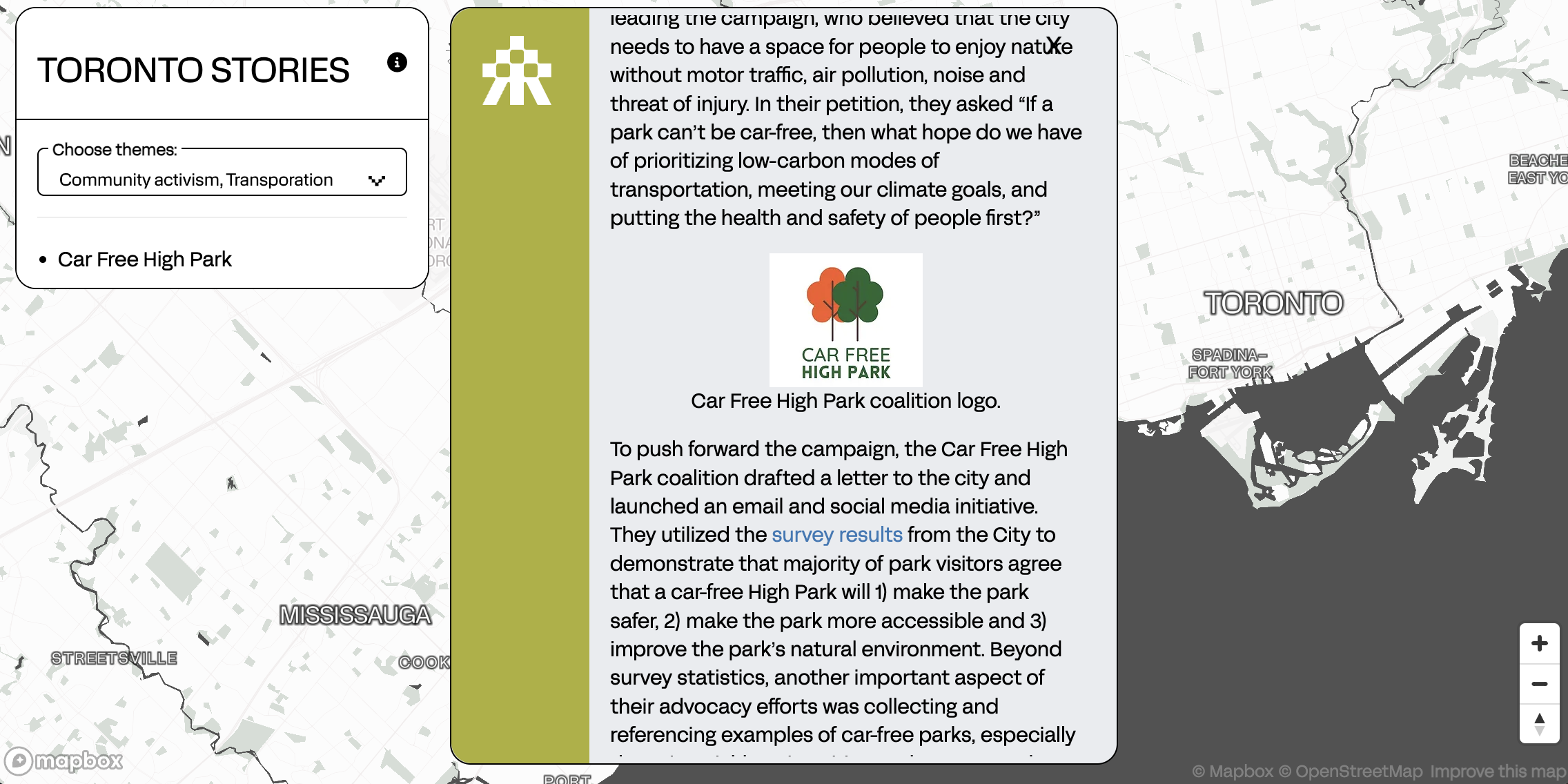}
  \caption
  {Toronto Stories module on \textit{Curbcut Toronto}. This module shows a map of place-based stories and advocacy campaigns happening around Toronto.}
  \Description{A screenshot of Curbcut Toronto's Toronto Stories module. The web page shown in the picture includes a header saying Toronto Stories, a drop-down menu titled chosen themes and the title of the story displayed in the middle, Car Free High Park. The story on display is a text-based document around the campaign of making high park car free on weekends. The story shows an image of the logo of the local campaign with two trees one orange and one green overlaying each other and four words in green below the trees saying car free high park. A map of the Toronto is being presented as the background behind the title section and the story section of the web page.}
  \label{fig:toronto-stories}
\end{figure*}

Beyond technical support for accessing and using far data, advocates see more importance in recognizing diverse ways of understanding what counts as data and what counts as meaningful data. For instance, even though \textit{Curbcut Toronto} is heavily focused on presenting quantitative data, many advocates highlighted the value of the \textit{Toronto Stories} module (Fig \ref{fig:toronto-stories}). P14 appreciated how the module adds “a human dimension” to the sometimes “cold” and “not inviting” data (DF3). Presenting stories with writings, photographs and art works beyond quantitative far data empowers community advocates to embrace diverse ways of knowing and storytelling. P17 also emphasized including and acknowledging data that are often ignored by institutionally supported open data programs. They pointed to the data and maps on the lost rivers of Toronto \cite{lost_rivers_lost_2024} which \textit{“influence a lot of things in our neighborhoods”}. Mapping these waterways centers Indigenous ways of knowing and understanding the city as a part of nature rather than apart from it (DF5). Although most urban data are associated with scientific ways of knowing associated with the census system, community advocates see much value in presenting and prioritizing local and Indigenous ways of knowing. 

The pathway of aligning data exploration with diverse ways of storytelling emphasizes both making far data nearer and bringing various near data from the periphery to the center. Bridging technical gaps help bring far data nearer to community advocates, making data more accessible and usable and offering more opportunities for community advocates to examine issues or biases embedded in far data. Through bringing far data nearer, data intermediaries can work together with community advocates to  examine power (DF1) and challenge power (DF2) around the shared concerns that are dominated by unequal power structures. At the same time, obtaining more space for marginalized ways of knowing is especially important for advocates. They push for incorporating historically marginalized near data used by Indigenous communities for understanding cities as well as storytelling techniques other than scientific analysis. Advocates' existing ways of working with data are reflected in data feminism principles of elevating emotion and embodiment (DF3) and embracing pluralism (DF5). They envision data intermediaries that align with and empower their diverse ways of storytelling, valuing multiple forms of knowledge, synthesizing varied perspectives through affording diverse ways of exploring and visualizing data. 

\subsection{Pathway 2: Communicate context and uncertainties}
While urban data is usually being presented in an authoritative narrative of accurate depictions of cities \cite{mattern2015, kitchin_knowing_2015}, community advocates are highly experienced in recognizing the limitations of urban data. Therefore, advocates emphasized that it is important for data intermediaries to communicate contexts and uncertainties behind data in order to open up spaces for dynamic discussion to imagine urban futures.

\subsubsection{Communicate context of far data} Nearly all participants recognized limitations of quantitative urban data through their various experiences of working with data for advocacy. To work with and around these limitations, participants emphasized the importance of questioning context when making sense of data. P11 highlighted the importance of questioning power and diversity behind data: \textit{“I think that data explorations are really important to talk about power and diversity of areas… It's often the most powerful group of people doing something that impacts others”} (DF6, DF1). P13 recognized that quantitative data only offers partial views, and they discussed the importance of incorporating qualitative data to provide contexts: \textit{“It's really important to have both [quantitative and qualitative data] for context, because numbers can't capture the entire impact that we're having…There are a lot of things that are not quantifiable”} (DF6, DF5). P5 and P1 emphasized the importance of \textit{“deliberately”} designing their survey questions for data collection to go beyond numbers by including questions focused on the \textit{“why”} to provide better context. P5 stated, \textit{“we chose this [survey question] because we really want to know why, what prevented people from biking.”}

Urban indicators, usually found in maps created by data intermediaries, add another layer of complexity to community advocacy work. Nearly all advocates were aware of the limitations and critiques related to urban indices. For example, when interacting with \textit{Curbcut Toronto}, many participants (P1,  P6, P7, P9, P11) critiqued the lack of context associated with the bikeway comfort and safety index. For example, P9 challenged the index through referring to their own lived experience, \textit{“this area probably does have the most by quantity bike lanes and bike infrastructure. But I also feel that this is an area that I don't like to bike to work because it doesn't feel safe, because it's so dense and there is so much traffic”} (DF3). P7 also commented on the nuances in feeling comfortable, \textit{“I wonder what their context of feeling comfortable is. There are individuals that think just because there's paint on the ground that you've got a bike lane. That's perfectly fine, but are you gonna let an 8 year old ride down a painted line or would you rather be in an actual protected piece of infrastructure”} (DF6, DF4). P9 posed several questions on the contextual information that they hoped to understand, \textit{“how is this data collected? Is this a survey? I wonder if this is about the literal number of infrastructure? Especially with a metric like this, if it's based on people's feelings of safety or some sort of audit or something.”} 

These experiences and expertise when sifting through data, questioning context, asking questions to delve into the context beyond data are transferred to their visions for data intermediaries’ work. Before P11 opened up \textit{Curbcut Toronto}, they commented: \textit{“before I even click into this, my biggest thought is: does it give folks data with enough context, which is a hard thing to automate. But that's always my thing because anyone can throw out a number, but sometimes they have ridiculous numbers”} (DF6). P11 added that when they use open data from different departments of the government, the discrepancies and various ways in presenting the same information, such as budget, constantly remind them that \textit{“data isn't just bulletproof infallible information.”} Other participants also recognized the dynamic nature of data, commenting that \textit{“nothing is perfect and everything is ever evolving”} (P15) and \textit{“we have to view it [data] as something that's living because it's changing all the time. People's experiences are changing all the time, and so that is never going to be stagnant, even though we might want it desperately to be”} (P16), and so data intermediaries should be mindful in acknowledging the dynamic nature and partiality of data.

\subsubsection{Reflect on intentions and values} Beyond acknowledging uncertainties behind data, community advocates also emphasize the importance of communicating the intentions and motivations behind data intermediaries' work. When presenting data through visualizations, P11 was aware of how different people can easily have different interpretations (DF6). They emphasized that providing context through complementary texts or blog posts is crucial for communicating their intent and stories with clarity, and explained:  \textit{“so the thing with data visualization is that I'm very careful not to ever present it on its own. Everyone learns in different types of ways and takes information in different types of ways. So I would never just send you this link and not tell you anything about it…I would embed this in a post about why you would want to look this up.”} P17 urges data intermediaries to be explicit about their positionalities, values, motivations behind their data works and why they think presenting a certain type of data is important to their values (DF6, DF7). P17 shared how different people might have different definitions on what urban sustainability means to them including \textit{“financial sustainability, environmental sustainability, or working towards a sustainable earth, a sustainable earth with humans.” Therefore, they appreciate when data intermediaries “name very clearly what [they] are aiming for.” }

Expressing uncertainties in data and processes of data analysis might be daunting for data professionals because the field has been conventionally driven by ideals of objectivity and neutrality. However, data are never raw or neutral \cite{gitelman_raw_2013}, and community advocates are very well aware of that. Pursuing the seemingly objective approaches of data science undermines the quality and effectiveness of both far data as well as near data. Being explicit of the limitations of the usually `objective' far data is a commitment to data feminism’s principle of rethinking binaries and hierarchies (DF4). Clarifying the positionalities and intentions of data intermediaries support the recognition and use of near data. When data intermediaries surface motivation of far data work, they not only provide future data users an important aspect of the context, but also acknowledge that far data, similarly to near data, are always coming from feeling bodies in this world, reflecting their values and situated knowledges \cite{haraway1988}. This supports working towards data feminism's principle of elevating emotion and embodiment (DF3), making labor visible (DF7), and embracing pluralism (DF5). Advocates’ envisioned pathway for communicating uncertainties and providing context is a crucial step towards the commitment to co-liberation \cite{DIgnazio2020} for both data intermediaries as data facilitators as well as other stakeholders relying on their work as voices that challenge dominating narratives and reimagine future cities.

\subsection{Pathway 3: Decenter artifacts for building relationships}
While community advocates see value in data intermediaries designing tools that support advocacy work, they highlighted the relational nature of their work and emphasized the importance of relationships beyond transactional interactions through technology artifacts. This theme echoes previous works looking into civic tech and community groups’ roles of capacities building beyond the mere production of end products \cite{mann2018,McCordBecker2023,walker2016}. Through fostering long term connections, data intermediaries and community advocates can collectively gain more civic capacity to connect near and far data for understanding urban issues and to build solidarity among different values in shaping urban futures.

\subsubsection{Synergize data work through relationship building} Most participants (P1, P3, P4, P5, P6, P7, P9, P10, P11, P12, P14, P16, P17) shed light on the relational nature of advocacy and how the values of prioritizing relationship building shaped their visions for data intermediaries. For some, building networks with and for community members is a key focus of their work (P1, P9, P10, P12). Advocates also shared that it is common to see people involved in multiple community groups, which opens up opportunities for working together and influencing each other. P11 emphasized that \textit{“I don't work in isolation and this ends up dictating what I'm working on.”} P7 summarized that \textit{“many advocates are wearing multiple hats with multiple lenses”} and work together towards the goals that \textit{“synergize with each other.”}

Advocates recognize that data is only a starting point of their work. This recognition often allows them to build more local capacity for better access to far data and for discovering valuable near data. When obtaining and analyzing urban data, community groups prioritize values of care \cite{Boone2023, Bopp2017}, building symbiotic relationships that support each other in resource-constrained situations. P6, P8, P13 and P14 all mentioned the importance of relying on their existing relationships to access and make sense of far data. P6 explained that they were able to access data related to how people use a local helpline for social services \textit{“only because we had that connection”} and \textit{“their GIS app helped us create the picture we want.”} Similarly, P11 shared that as they became familiarized with a data space, they were happy to help other groups to navigate. Advocates also described how they support other advocates and community groups through what P1 called a \textit{“symbiotic relationship.”} P1 described how they got into advocacy work through the help of other advocates’ networks and commented that \textit{“my connections with all those others have been strengthened and my goal is always to support the organization and then together it’s a symbiotic relationship. I get better from their activities, they benefit from my activity.”}

Advocates' focus on relationship building also fostered a more collective effort in empowering the use of near data. P1 started their initiative on collecting urban noise data and was able to make visualizations and build a website for the project through connecting with other people and sharing their experiences at local events, where collaborations emerged. P2 shared their experience working with a group of people to push forward a campaign on local park access, where they together collected thousands of local survey responses about people’s experiences in parks. P3 collaborated with journalists to write articles that make visible their lived experiences navigating urban life after a brain injury. This amplified voices of those experiencing hurdles related to physical recovery, mental health, and access to affordable housing. These advocacy efforts in collecting and telling stories leveraging near data were achieved through advocates’ constant effort in building long term and trusting relationships through local events, sharing back with their communities and embracing new relationships that arise through their work (DF2, DF5, DF7). The dynamic, organic relationship among advocates was summarized by P7 as \textit{“it [advocacy work] may start with me, but I'm only there for a little bit before I'm bringing in a data visualization person, or a teacher from a school, or a community member that's been in the neighborhood for 25 years, or the politician, and maybe I get to step back and I let others go forward.”}

\subsubsection{Build relationships with and for community advocacy groups} Community advocates’ attention to relationships beyond mere transactions were again reflected when they interacted with \textit{Curbcut Toronto}. They emphasized the vision of building long term relationships with data intermediaries and expanding their network through data intermediaries’ work. Several participants mentioned that they would not trust a data intermediary tool without knowing the team behind it (DF7). P6 commented that \textit{“I would trust it if it was offered to my organization through outreach, or I heard about it through a workshop or webinar.”} Similarly P7 mentioned that data intermediary tools need ambassadors and advocates \textit{“who can speak to the clarity of this tool”} or \textit{“launch an AMA, ask me anything.”} Many participants (P10, P14, P15, P17) also shared that the information provided by a data intermediary might not be the most important aspect, but a starting point for conversations and discussions for building connections among different stakeholders. P10 commented that \textit{“we have team members who are still in school and we have team members who are seasoned professionals”} and the tool could help \textit{“open up a lot of ways of understanding for the team so we can share our thoughts and have a mutual connection”} (DF5). P17 stressed the importance of data intermediaries to present other group’s ongoing effort in leveraging near data, so that groups will be able to learn from each other and work in solidarity (DF7): \textit{“what I would do personally is if I saw a case study dealing with the same issue over here, I would give them a call…[Data intermediary tool] doesn't need to provide all the information, it just needs to provide a story so that I can go and talk to them myself.”}

Advocates found data intermediary’s work to be more valuable if they collaborate with existing local social infrastructures and civic groups. To better support ongoing community work, P1 suggested data intermediaries to connect with communities such as civic tech groups and local schools, explaining that \textit{“it is not necessarily presenting a tool for people to use for their own purposes, but also educating people on their surroundings.”} P5 discussed how data intermediaries could bridge the gap for high school students who are starting to learn about urban information. P7 discussed connecting with an even broader audience and emphasized that it is important to \textit{“get the regular public to be aware of the things around them and have easy access to all of these data pieces…don't shy away from being able to have elementary schools gain access to something like this.”} P3 explained how they benefited from affordable classes offered by their local libraries and are advocating for more: \textit{“So let's say if I was taking a class on Google tools, or Photoshop or Windows 11, we should have the same thing with the information where we teach our community members at a minimum cost to learn how to get to the data.”} P7, giving back to their community through creating an interactive map on community resources for their local library, commented \textit{“the library is being really active in being a community hub as a piece of social infrastructure. So that's my little give back with all of these maps that I'm creating.”}

When working with data, advocates pointed out that data is not the purpose but relationship and capacity building is much more important. P1’s eventual goal of their data work is going beyond data \textit{“to push it out to the public to educate, to engage, to make aware,”} and so they are learning press releases and also writing a blog on the Toronto Open Data Portal. For P4’s group, the primary goal of their data collection project \textit{“was not the data, it was the conversation. So we were trying to get people to think about their home energy use.”} Similarly, P11 involved data in their work because \textit{“it empowers people to speak with their friends, their family and their colleagues.”} P17 emphasized that the crux of advocacy work lays beyond data in finding a strategy that can move forward even when there are divergent views. Pursuing this pathway allows data intermediaries to follow data feminism’s call for looking at data from an embodied perspective, putting more attention to the people who work on the data and use the data, and making that labor visible (DF3, DF7). Advocates share an understanding that data tools are only starting points of supporting ongoing efforts to shape urban futures beyond dominant imaginaries. Only when local networks are formed will there be more capacity for connecting near data and far data, embracing diverse perspectives, and finding solidarity in differences to enact change.



\section{Discussion}
\subsection{Actions for putting data feminism into practice}

Data intermediaries are never neutral. Previous works in HCI and CSCW have illustrated how community advocates enact data feminism in their data advocacy work. For instance, Darian et al. \cite{Darian2023} presented how nonprofit advocacy groups enact data feminism principles when reappropriating data across various functions from the margins of the data economy. Paudel and Soden \cite{paudel_reimagining_2023} used the data feminism principles to evaluate open data platforms and highlighted the gaps in how these platforms lack consideration in values from data feminism, such as making labor visible, engaging with affective aspects of disasters, and challenging power. Tran et al. \cite{tran_situating_2024} studied local housing activist organization's data work through the lens of data feminism and proposed design implications when situating data for grassroot organizers. The work of counterdata is also surfaced by D'Ignazio \cite{D'Ignazio2022} as a central practice for data activism, especially when data intermediaries fail to produce marginalized information and reproduce oppressive narratives. Extending this growing body of work, we reflect on how community advocates' vision for data intermediaries surface data feminism principles, and we provide concrete actions for data intermediaries to put data feminism into practice.


\begin{figure*}[h]
  \centering
  \includegraphics[width=\textwidth]{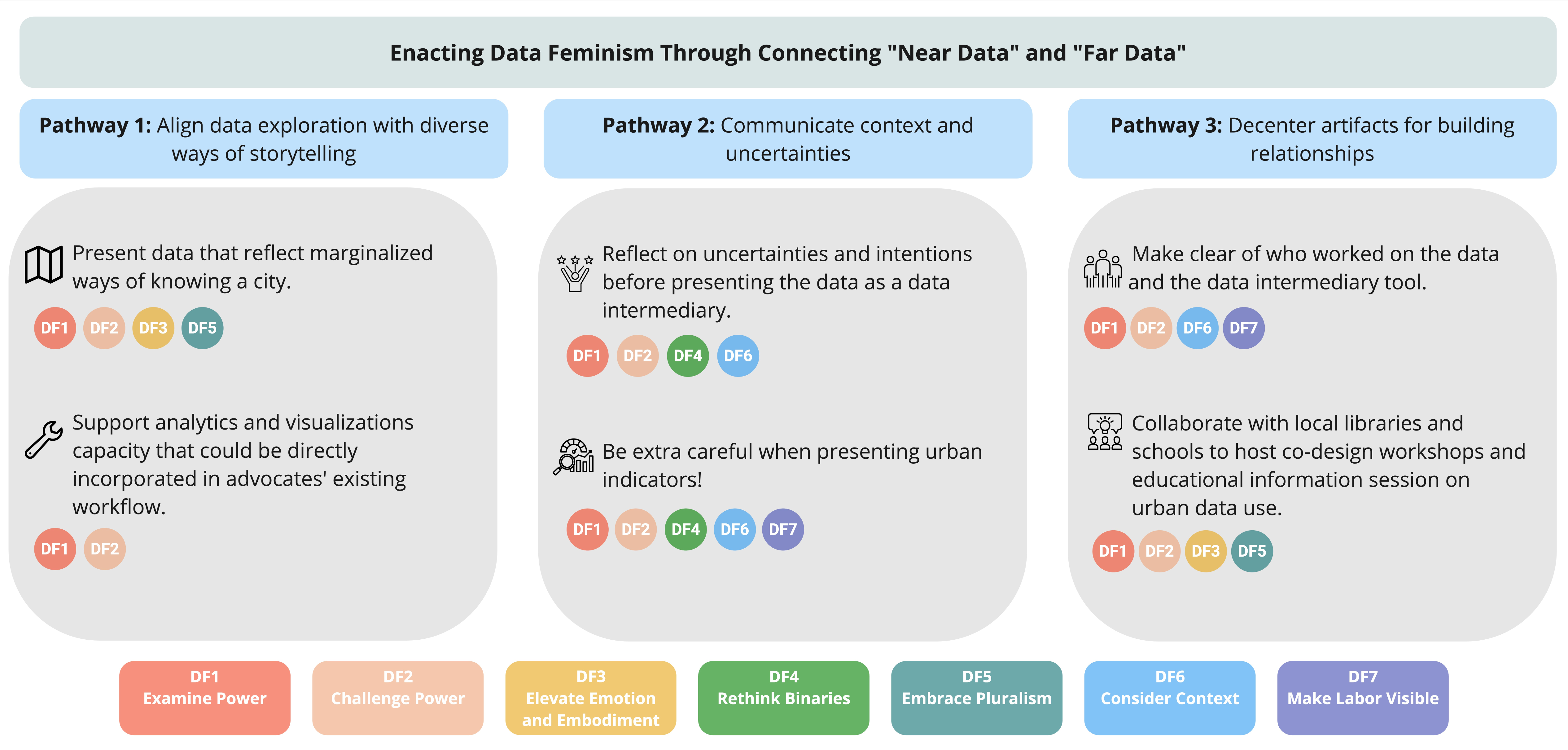}
  \caption
  {Starting points and actionable items suggested by community advocates for data intermediaries to connect near data and far data. Each recommendation was raised by multiple community advocates and speaks to data feminism principles.}
  \Description{}
  \label{fig:actionable_items}
\end{figure*}

 Throughout the results section, we illustrated how our interviewees' ways of working with data and their visions for data intermediaries reflect principles of data feminism, labeling their quotes with DF1 to DF7. We distill advocates' insights and perspectives into Figure \ref{fig:actionable_items} to offer data intermediaries several actionable next steps for connecting near data and far data along each of the three pathways.
 
 Each recommendation is tagged with the related data feminism principles. We argue that following these recommendations can help data intermediaries enact data feminism and support those who rely on their work to further enact data feminism. For instance, intermediaries should present data that reflect marginalized ways of knowing a city, such as the lost rivers of Toronto \cite{lost_rivers_lost_2024}. By incorporating marginalized experiential ways of knowing, data intermediaries can examine power (DF1) to challenge power (DF2). Their work in turn can elevate emotion and embodiment often neglected by data-centric tools (DF3) and can serve as a medium for others to embrace and create plural narratives of urban spaces (DF5). They should be careful when presenting urban indicators, decomposing the messy meaning and work behind the numbers. These actions help data intermediaries to rethink binaries (DF4) often introduced in urban indicators, consider the context (DF6) of why and how these indicators were created, and make those who worked on defining and using these indicators visible to the users (DF7), whether the users engage with these indicators or refuse to be constrained by them. These actionable items serve as concrete starting points for those data intermediaries who are ready to enact data feminism to shape urban futures that center justice, diversity and inclusivity.

\subsection{Data intermediaries' roles in pursuing ``The Right to the City''}

As Currie et al. \cite{Currie2022} put forward, data work is central for pursuing the Right to the City in the datafied city today. We build on this argument and argue that data intermediaries play crucial roles in pursuing the Right to the City.

Data intermediaries work is never just making it easy for more people to access data. Every design decision -- what data to include and exclude, how to frame explanations, and how the interface guides user interaction -- is political. As Loukissas argues, ``interfaces recontextualize data'' \citep[p.125]{loukissas_all_2019}. Data intermediaries create new contexts for data that shape how users engage with the information, what ways of knowing are made available, how users connect data into their own knowledge systems, and how they will navigate their right to the city. When data intermediaries recognize the distance between their work and the data they work with, they become more attuned to the data’s original contexts and how their processing and design choices reshape it.

Data intermediaries’ work is also inextricably linked to the Right to the City through addressing data justice concerns \cite{Currie2022}. There are different approaches to data justice \cite{han_psychopolitics_2017, eubanks_automating_2018, dencik_conceptual_2018, oneil_weapons_2016, taylor_what_2017, noble_algorithms_2018}. Here, we use Taylor’s framework \cite{taylor_what_2017} to illustrate data intermediaries roles, as it clearly recognizes both the potential and limitations of data technology. Under the framework, there are three pillars of data justice: visibility refers to the right to be visible and invisible; engagement refers to the freedom of accessing and the freedom to not use particular technologies; non-discrimination emphasizes challenging bias and the freedom to be not discriminated against. The vision of connecting near data and far data is closely linked to all pillars of data justice. As discussed in Section \ref{sec:vision}, advocates pointed out how connecting near data and far data allowed them to make their lived experience more visible to the public. At the same time, advocates emphasized how data intermediaries’ work of aggregating and merging data serves as key process to protect privacy for those who wish to remain anonymous. Connecting near data and far data also ease the daunting process of making sense of far data, supporting those who hope to engage. Decentering artifact for relationship building allows data intermediaries to foster relationships beyond mere transactional interaction \cite{McCordBecker2023}. When conversation is opened up and local networks are fostered, community advocates will be empowered to decide when not to use a technology because they can achieve their advocacy goals through relational ways of skills and data sharing. Lastly, connecting near and far data supports users to notice what is being discriminated against and what is missing. Rather than elevating far data to the point where it marginalizes and invalidates lived experience, the vision guides data intermediaries to support users in appreciating diverse types of data, examining whose data is represented and how data is presented. For many advocates, lived experiences are in fact crucial both for making sense of that far data and for pinpointing the biases and uncertainties embedded in those data. When near data, such as shared experiences, bring advocates together, these collective efforts will have a greater impact on pursuing the Right to the City.

\subsection{Locating “near data” and “far data”}
Near and far are not static qualities of data but dynamic, fluid and relational qualities, dependent on who is working with the data, how it is presented and passed down by data intermediaries across the urban data ecosystem, and to whom it is presented. When interacting with data, do we know the people who created the data? Do we know how the data are collected, processed, and presented? What are the values and intentions of the people who selected the data and worked on them? How were these values baked into the data when being collected and processed? \cite{shilton_how_2014}

The quality of near and far can be influenced by geographical and temporal, but most importantly social and socio-technical dimensions. Drawing from social and environmental psychology research, Bhardwaj et al. \cite{bhardwaj_limits_2024} argued that psychological distance can be understood in four dimensions (social, temporal, geographical, and hypothetical) and that these dimensions present important factors for data visualization designers to consider as they affect how people perceive, negotiate, reason and decide \cite{trope_construal-level_2010,rim_what_2013}. Similarly, we argue that the quality of "near” and “far” is especially important for data intermediaries to consider when working with data and presenting data. Data are considered near when they arise in close proximity: for example, from our own experience of biking in our own city in the recent past. In contrast, projections about future urban heat islands and their effects on bikeway comfort may be geographically and socially close but are temporally removed and hypothetical. How advocates relate to a data set presenting such information, how they work with it, and how they decide to trust it is contingent on this distance. Anticipating that relationship can help data intermediaries design the appropriate context documentation, including heat models and evaluations underpinning a comfort index. 

Many advocates collect near data, including the urban noise dataset for P1, community garden mapping for P4, as well as biking experience documentation for P7. These are near data because the advocates know who is collecting the data, how the data is collected and why the data is collected. To someone who comes across these datasets online, they may not be: it depends on how the data intermediaries re-present these data. With the right documentation and explanation, the data can become closer to those encountering them. That is a central role and value of data intermediaries. When data intermediaries provide data access and analytics support to other stakeholders in the urban data ecosystem, they should reflect on their distance to the data. Their design decisions will impact how others experience and understand the data as near or far.

As Loukissas argues in “All Data are Local” \cite{loukissas_all_2019}, data workers and users should learn to analyze data settings rather than data sets, attending to the places, institutions, processes and people that shape the data. This is why near data is extremely valuable. Designs of community-oriented advocacy platforms, such as \textit{Smell Pittsburgh}, are often most effective when they seamlessly embed individually collected data, near to the users who submit them, within the geographic and temporal context of community advocacy \cite{hsu_smell_2019}. By allowing community members to recognize others' data instances within the community context, the platform as a whole becomes an effective advocacy tool and is able to present the sum of the near data as a far-reaching dataset. We contend that reflecting on the nearness and farness of data offers a concrete approach for data intermediaries to attend to data settings.

When we emphasize the importance of near data, we are not dismissing far data, but in fact, far data offer fresh perspectives especially for those engaging from afar. Historical data, for instance, allow us to uncover stories our ancestors experienced and reveal temporal trends. Spatially distant data help us recognize differences and facilitate comparisons. However, we need to be careful with abstractions associated with far data. Far data risks being regarded as singular fact presented in a "view from nowhere" that, as feminist scholars point out, is an illusion \cite{haraway1988, DIgnazio2020}. In Seeing Like a State \cite{scott_seeing_1999}, Scott examines how scientific rationalism is applied to govern people through simplification, raising concerns about decision-making processes that reduce everyday experiences to calculable abstractions. In CSCW and HCI, Dourish \cite{dourish_seeing_2007} argues that the development of information systems is strongly related to systems rationalism underlying the phenomena that Scott analyzes. More recently, Wong et al. \cite{wong_seeing_2023} also shows how Scott’s concept of simplification to govern is reflected in AI ethics toolkits. In the context of data intermediaries, presenting far data through abstraction and aggregation of geographic or temporal scale is not necessarily bad: data users are not expecting to see every tree in a city throughout history labeled on a map. However, data intermediaries should reflect on their distance to the data, their processes of abstraction, the locally situated practices and near data that are abstracted away, and prioritize design decisions that enable users to also reflect, examine and challenge the data provided.


\subsection{Tensions of connecting “near data” and “far data”} 
We recognize that connecting near data and far data is a complicated and messy process. Here we illustrate two tensions that could arise and call for research into how data intermediaries and community advocates can work together to navigate these tensions.

We recognize that urban governance prioritizes conventional data science and technical analysis, which leads to the tension between codified statistical analysis and advocates’ experiential ways of knowing and storytelling. As P9 noted, high-level statistical analysis, while invaluable, may not always be the most effective tool for advocacy work: \textit{“I simplified it [regression model] into like ‘there is an association between these variables’... because, for our audience, it's not really that easy to interpret.”} Neff et al. \cite{neff_critique_2017} articulated that “stories occur before data production, during production, and are used in exchange to give data meaning across communities with different expertise, cultures, and practices.” This perspective highlights the significance of storytelling at every phase of data work and emphasizes the value of embracing diverse cultures and practices when working with data. This tension has been pointed out by many scholars in feminist HCI, critical data studies and adjacent fields \cite{dignazioklein2021,aragon_developing_2016,gitelman_raw_2013, erete2016}. It intensifies in the context of designing data exploration tools for urban sustainability advocacy. We invite researchers to explore in the context of data intermediaries: how can data intermediaries work with advocates to prioritize experiential ways of knowing over codified statistical analysis in practice? How would this differ across the spectrum of data intermediaries work and various types of advocacy work?

We also recognize the tension between community advocates’ hope for an \textit{“impartial third party”} to provide data and the impossibility for any data intermediary to be genuinely impartial. During our interview, P1 explained the push back they got when advocating for bike lanes in local neighborhoods,  \textit{“car people tend to look at our information with a grain of salt. They're like, you guys are biased in this, in that. So having impartial third party data to show is hugely important.”} P3, as someone who is distrustful of many sources of data, emphasized that \textit{“we need access to accurate data where a third party is collecting the data.”} Even though there is a need and tendency for data projects to present themselves as objective and authoritative \textit{"third parties"} \cite{dignazioklein2021}, it is crucial for data intermediaries to make clear their positionalities, intentions and values that influenced the work that they are doing, to work towards co-liberation, the underlying commitment of data feminism \cite{dignazioklein2021}. To continuously work towards liberating data science from the dominant view of objectivity and authority, key questions arise for the HCI and CSCW community: how can data intermediaries better communicate intentions and uncertainties in data and data visualization? How can advocates and data intermediaries work together to reflect on contexts and uncertainties from multiple perspectives? How can reflexivity become a key component of data work? \cite{miceli_documenting_2021} We see value in keep asking these questions and invite more HCI researchers joining us to work towards datafied urban futures that center social justice and inclusivity.

\section{Conclusion}
To quote Neff et al., data intermediaries should strive to make data "less terrible" \citep[p.88]{neff_critique_2017}, not more, and this requires data intermediaries to “think with, not for or against” \citep[p.93]{neff_critique_2017} community advocates. Community advocates have rich lived experience and unique perspectives to offer those of us working with and studying data. This research explores community advocates’ visions for data intermediaries, a crucial stakeholder in the urban data ecosystem facilitating interactions amongst almost all stakeholders. Through conducting interviews with 17 community advocates working with 23 different community groups, we surface the important quality of near and far data to be considered in data intermediaries' work. We contribute a description and reflection on community advocates’ vision for data intermediaries, connecting near data and far data, as well as three pathways for pursuing this vision: align ways of data exploration with diverse ways of storytelling, communicate context and uncertainties, and decenter artifacts for relationship building. We illustrate how following the pathways allows data intermediaries to bring data feminism principles into the specific context of their work. To conclude, we propose several actionable next steps for data intermediaries, discuss on how these pathways contribute to the pursuit of ``the Right to City'' in the datafied city today, and reflect on tensions of working within the urban data ecosystem. We reinforce the importance of questions around the role, intention, and impact of data intermediaries. With relational ways of working that center justice and pluralism, data intermediaries can collectively work together with other stakeholders to pursue the Right to the City: a change that is more than legal, technical or procedural, but a “political capability -- a configuration of civic life that can enable and nurture solidarity as grounds for struggle.” \citep[p.XVi]{dencik_foreword_2022}

\section{Acknowledgements}
This research was partially supported by NSERC through RGPIN-2016-06640, the Canada Foundation for Innovation, and the Ontario Research Fund.

\bibliographystyle{ACM-Reference-Format}
\bibliography{sample-base}

\appendix









\end{document}